\documentclass[twocolumn,aps,prl,floatfix]{revtex4-1}
\usepackage[utf8]{inputenc}
\usepackage{graphicx}
\usepackage{dcolumn}
\usepackage{times}
\usepackage[breaklinks]{hyperref}
\usepackage[normalem]{ulem}
\usepackage{braket}
\usepackage{mathrsfs}

\usepackage{amssymb} 
\usepackage{amsmath}
\usepackage{amsthm}

\renewcommand*{\HyperDestNameFilter}[1]{\jobname-#1}

\renewcommand{\vec}[1]{\mathbf{#1}}

\newcommand{\hatn}{{\hat{\vec n}}}

\newcommand{\sgn}{\text{sgn}}

\newcommand{\Gret}{G^{\text{R}}}

\newcommand{\com}{M_{\text{com}}}
\newcommand{\tom}{M_{\text{tom}}}
\newcommand{\Om}{\mathbf{M}_\text{om}}
\newcommand{\Com}{\mathbf{M}_{\text{com}}}
\newcommand{\Tom}{\mathbf{M}_{\text{tom}}}

\newcommand{\dd}{\text{d}}
\newcommand{\tr}{\text{tr}}

\newcommand{\xc}{\Delta_{\text{xc}}}
\newcommand{\soi}{\alpha_{\text{R}}}
\newcommand{\mass}{m_{\text{e}}}

\begin{document}

%\preprint{APS/123-QED}

\setcounter{secnumdepth}{2} 

\title{Chiral and Topological Orbital Magnetism of Spin Textures}

\author{Fabian R. Lux}
\email{f.lux@fz-juelich.de}

\author{Frank Freimuth}
\author{Stefan Bl\"ugel}
\author{Yuriy Mokrousov}

\affiliation{Peter Gr\"unberg Institut and Institute for Advanced Simulation,
Forschungszentrum J\"ulich and JARA, 52425 J\"ulich, Germany}

\date{\today}

\begin{abstract}
%Fascinating properties of chiral topological magnetic textures such as domain walls and skyrmions have given rise to a plethora of novel concepts in modern spintronics.  
%sIn this work, 
Using a semiclassical Green's function formalism, we discover
the emergence of chiral and topological orbital magnetism in two-dimensional chiral spin textures by explicitly finding the corrections to the orbital magnetization, proportional to the powers of the gradients of the texture. We show that in the absence of spin-orbit coupling, the resulting orbital moment can be understood as the electronic response to the emergent magnetic field associated with the real-space Berry curvature. By referring to the Rashba model, we demonstrate that by tuning the parameters of surface systems the engineering of emergent orbital magnetism in spin textures can pave the way to novel concepts in orbitronics.
\end{abstract}

%\pacs{75.10.Lp, 03.65.Vf, 71.15.Mb, 71.20.Lp, 73.43.-f}
\maketitle

{\itshape Introduction.}~The importance of chiral magnetic structures such as domain walls and skyrmions is ever growing due to their role in the formulation of advanced concepts in spintronics, accompanied by frequent discoveries of novel effects hinging on the finite chirality of these particle-like textures~\cite{Fert2013}. 
In recent years significant advances have been made in the reliable
detection of chiral textures and in their efficient manipulation by external perturbations~\cite{Jiang2015}.
For the case of skyrmions, the palette of emergent topological phenomena they give rise to~\cite{Nagaosa2013}
was shown to root predominantly in the ``emergent" magnetic field, intrinsically generated by the non-trivial real-space distribution of the orientation of the magnetization, $\hat{\mathbf{n}}(x,y)$:
\begin{equation}
B_{\text{eff}}^z = \frac{\hbar}{2	e}\ \hat{\mathbf{n}} \cdot \left( \frac{\partial \hat{\mathbf{n}}}{\partial x} \times \frac{\partial \hat{\mathbf{n}}}{\partial y} \right).
\label{eq:Beff}
\end{equation}
The integrated flux of this field gives rise to a quantized integer topological charge of a skyrmion, $N_\text{sk}$, affecting its dynamical properties and resulting~e.g.\ in an enhanced robustness with respect to scattering and fluctuations~\cite{Sampaio2013}.

Within an intuitively appealing scenario, the emergent magnetic field in chiral systems couples directly to the orbital degree of freedom
and provides an alternative mechanism for Hall effects, such as the topological Hall effect of skyrmions or the anomalous Hall effect in non-collinear 
magnets~\cite{Hoffmann2015, Kubler2014, Bruno2004}. 
Besides these well-known effects, the orbital response to $B_{\text{eff}}^z$ gives rise to a phenomenon which was coined
{\it topological orbital magnetization} (TOM)~\cite{Hoffmann2015}, and which has not yet been explored to the same extent. 
Indeed, the emergence
of orbital magnetism that does not rely on spin-orbit interaction in strongly frustrated systems and small-size skyrmions has been shown from first-principles and tight-binding calculations~\cite{Hanke2016,Hanke2017,Shindou2001, Dias2016}. 
The novel chirality-driven ``topological" channel for the orbital magnetism is a remarkable, attractive effect, since it provides an
additional control on the handedness of the underlying texture, may exhibit enhanced robustness~\cite{Dias2016}, and generally paves the way for non-trivial
spin textures into the realm of orbitronics with the vision of addressing and operating with orbital degrees of freedom of electrons rather
than their spin~\cite{Hanke2017}.  
However, at the current stage, the physics of the TOM is very poorly understood and the rigorous framework for accessing orbital magnetism in chiral spin textures, imperative for our ability to engineer 
and utilize the orbital degree of freedom of chiral systems for the purposes of orbitronics, has been missing so far.   

In this Letter, by refering to the Green's function perturbation theory \cite{Onoda2006}, we put the orbital magnetism in chiral systems on firm quantum-mechanical ground presenting a rigorous theory for the emergence  of  orbital magnetism in non-collinear systems. By systematically tracing the orders of perturbation theory for chiral magnetic textures we distinguish corrections to the out-of-plane orbital magnetization $\Om  =  M (\hat{\mathbf{n}}) \mathbf{e}_z$ of a locally ferromagnetic system, appearing in higher orders of the gradients of the magnetization:
\begin{align}
 & \Com  =  M^\alpha_i (\hat{\mathbf{n}})( \partial_i  n_\alpha )\mathbf{e}_z
 \label{eq:com}
\\
& \Tom  =  M^{\alpha\beta}_{ij} (\hat{\mathbf{n}})( \partial_i  n_\alpha )( \partial_j  n_\beta )\mathbf{e}_z,
\label{eq:tom}
\end{align}
where $\partial_i = \partial / \partial x^i$. Here, and in the following discussion, summation over repeated indices is implied, with greek indices $\alpha,\beta \in \lbrace x,y,z \rbrace $ and latin indices $ i,j,k \in \lbrace x,y \rbrace$.

\begin{figure}[t]
 \centering
 \includegraphics[width=0.9\linewidth]{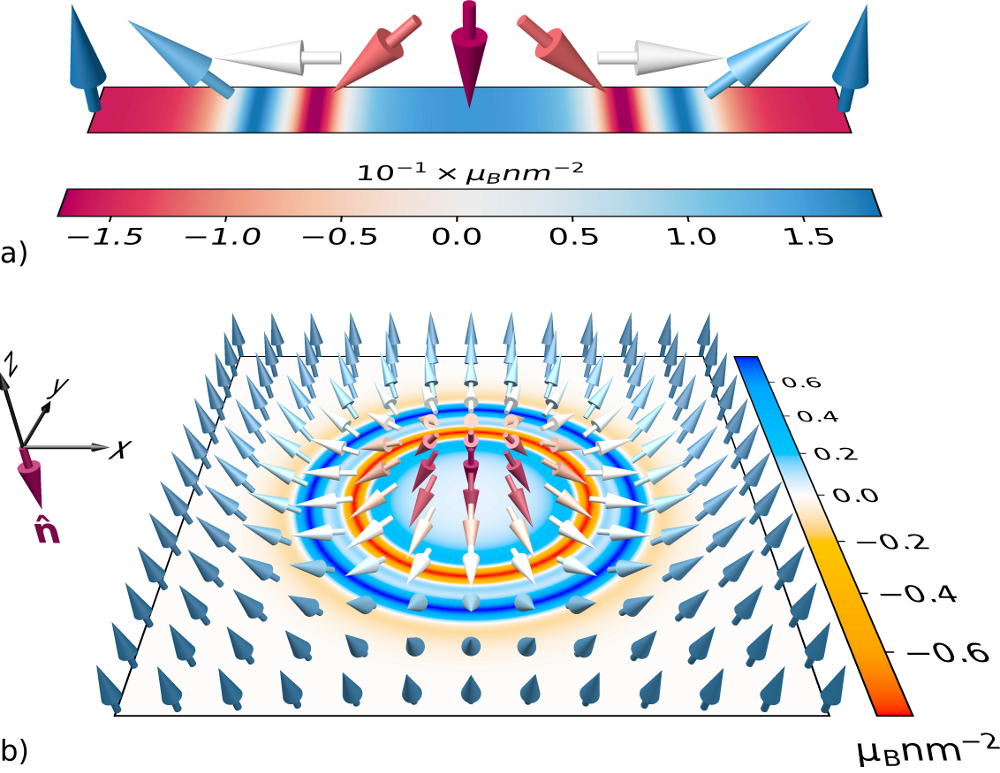}
 \label{fig:skyrmdens}
 \caption{In one-dimensional and two dimensional chiral topological spin textures, such as spin-spirals (a) or skyrmions (b), the non-trivial distribution of the magnetization $\hatn$ gives rise to (a) chiral orbital magnetization (COM), linear in the gradients of $\hatn$, Eq.~(\ref{eq:com}), and to (b) topological orbital magnetization (TOM), which is second order in the gradients of 
 $\hatn$,
 Eq.~(\ref{eq:tom}). Both COM and TOM can exhibit a complex distribution in real space, as illustrated in (a) and (b) respectively by explicit calculations for systems discussed in the text (with parameters 
 $\xc = 0.9$\,eV, $\hbar\soi = 2.0$\,\,eV$\mbox{\AA}^{-1}$, $k_\text{ B} T = 0.1$\,eV).
}
  \label{fig:localdens}
\end{figure}

In addition to the effect of TOM, appearing at the second order (Eq.~(\ref{eq:tom}) and Fig.~\ref{fig:localdens}(b)), we thereby propose a novel contribution to the orbital magnetization, which is linear in the chirality of the underlying texture (Eq.~(\ref{eq:com}) and Fig.~\ref{fig:localdens}(a)), and which we thus call the \itshape chiral orbital magnetization \normalfont  (COM). By explicitly referring to the
2D Rashba model, we numerically evaluate the magnitude and real-space behavior of the TOM and COM, finding that by tuning the parameters of surface and interfacial systems the orbital magnetism of domain walls and chiral skyrmions can be engineered in a desired way. Our findings open new vistas for exploiting the orbital magnetism in chiral magnetic systems, thereby launching the field of chiral orbitronics.

The expansion in magnetic field gradients is naturally achieved within the phase-space formulation of quantum mechanics, the Wigner representation~\cite{Onoda2006, Freimuth2013}. The key quantity in this approach is the retarded single-particle Green's function $\Gret$, implicitly given by the Hamiltonian $H$ via the Dyson equation
\begin{equation}
\left(
\epsilon
- H(X,\pi ) + i0^+
\right)\star \Gret(X,\pi) = \text{id} ,
\label{eq:dyson}
\end{equation}
%\todo{use $\text{id}$ for generality or simply $1$ for accessibility?}
where $X^\mu = (t,\mathbf{X})$ and $\pi^\mu = (\epsilon,\boldsymbol{\pi})$ are the four-vectors of position and canonical momentum, respectively. The latter of the two, in terms of the elementary charge $e>0$ and the electromagnetic vector potential $A$, is related to the zero-field momentum $p$ by the relation 
$
\pi_\mu (X,p) = p_\mu + e A_\mu(X) .
$
The $\star$-product, formally defined by the operator
%\begin{equation}
$\star \equiv \exp \left\lbrace\frac{i \hbar}{2}  \left(
[\overset{\leftarrow}{\partial}_{x^\mu}   , \overset{\rightarrow}{\partial}_{\pi_\mu} ]
- e F^{\mu\nu}\overset{\leftarrow}{\partial}_{\pi^\mu}
\overset{\rightarrow}{\partial}_{\pi^\nu}
 \right)   \right\rbrace$
%\end{equation}
of left- and right-acting derivatives $\tiny\overset{\leftrightarrow}{\partial}$, allows for an expansion of $\Gret$ in powers of $\hbar$, gradients of $\hatn$ and external electromagnetic fields, captured in a covariant way by the  field tensor $F^{\mu\nu} = \partial_{x_\mu} A^\nu - \partial_{x_\nu} A^\mu$~\cite{Onoda2006}.

In this work, we are after the orbital magnetization (OM) in $z$-direction. Given the grand canonical potential $\Omega$, the surface density of the orbital moment is given by~\cite{Zhu2012,Shi2007}
\begin{equation}
M(x) = - \partial_B \Braket{\Omega(x)} ,
\label{eq:magformula}
\end{equation}
which requires an expansion of $\Omega$ up to at least first order in the magnetic field $\mathbf{B} = B \vec{e}_z$ in the collinear case. In the limit of $T\to 0 $, the grand potential is asymptotically related to the Green's function $\Gret$ via
\begin{equation}
 \Braket{\Omega} \sim - \frac{1}{\pi}
 \ \Im\   \int \frac{\dd p} {(2 \pi \hbar)^2}   \ f(\epsilon)  (\epsilon - \mu) \ \tr\ \Gret (x,p),
 \label{eq:grandpot}
\end{equation}
where $\Im$ denotes the imaginary part, the  integral measure is defined as  $\dd p = \dd \epsilon\ \dd^2\mathbf{p}$, $f(\epsilon)$ represents the Fermi function $f(\epsilon) = (e^{\beta (\epsilon-\mu) } +1 )^{-1}$, $\mu$ is the  chemical potential and $\beta^{-1} =k_\text{B} T $. In our approach, deviations from the collinear theory enter the formalism as gradients of $\hatn$ and can be traced systematically in $\Gret$ and in $\Braket{\Omega}$, finally leading to Eq.~(\ref{eq:com}) and (\ref{eq:tom}). For details concerning the analytical and numerical strategy of devising this expansion in a diagrammatic way we refer to the supplement.

While our approach is very general, for the purposes of illustration and feasible numerical estimates, we restrict our further analysis to the two-dimensional Rashba model 
%in Wigner representation
\begin{equation}\label{rashba:ham}
H(X,\pi) = \frac{\pi^2}{2 \mass} + \soi ( \boldsymbol{\pi} \times \boldsymbol{\sigma} )_z + \xc\ \boldsymbol{\sigma} \cdot \hat{\mathbf{n}}(\mathbf{X}),
\end{equation}
where $\mass$ is the electron's (effective) mass, $\boldsymbol{\sigma}$ denotes the vector of Pauli matrices, $\soi$ is the Rashba spin-orbit coupling constant, and $\xc$ is the strength of the local exchange field. This model has been proven to be extremely fruitful in unravelling various phenomena in surface magnetism~\cite{Manchon2015}.

{\itshape Chiral Orbital Magnetization.}~When $\Omega$ in Eq.~(\ref{eq:grandpot}) is expanded in orders of $B$ and in orders of the
exchange field gradients, the first correction to OM appears at the second order in these perturbations. It is
the COM, Eq.~(\ref{eq:com}), and it is linear in the exchange field gradients.
To get a first insight into this novel effect, we consider an instructive example of a spin-spiral solution with wave-vector $q$ propagating in $x$-direction according to the form $\hat{\mathbf{n} } = (\sin(qx) ,0, \cos( q x)  )^T$, as depicted in Fig.~\ref{fig:localdens}(a).
For this N\'eel-type texture, one finds that up to $\mathcal{O}(\soi)$ the local $x$-dependent orbital moment is given by
\begin{equation}\label{com:model}
\frac{\com}{\cos \theta} =  \frac{e q \soi }{48 \pi} \sgn(\xc)\left( 1 - 3 \frac{\mu^2}{\xc^2} \right)  \Theta( | \xc | - | \mu | ), 
\end{equation}
where $\theta=qx$ is the angle between $\hat{\mathbf{n}}$ and the $z$-axis, and $\Theta$ is the Heaviside step-function. In the vicinity of $\soi = 0$, the magnitude of COM is thus proportional to the strength of spin-orbit interaction and vanishes in the limit of zero $\soi$. 

It is rewarding to understand this behavior in the language of gauge fields. To linear order in $\soi$, the Rashba Hamiltonian can be expressed as a perturbative correction to the canonical momentum
$\boldsymbol{\pi} \to \boldsymbol{\pi} + e \boldsymbol{\mathcal{A}}^R$, with 
$\boldsymbol{\mathcal{A}}^R =  m \soi \epsilon^{ijz} \vec{e}_i \sigma_j / e$.
For $|\soi| \ll |\xc|$ the spin polarization of the wavefunctions is only weakly altered away from $\hat{\mathbf{n}}$ and we can use the $SU(2)$ gauge field, defined by $\mathcal{U}^\dagger  (\boldsymbol{\sigma} \cdot \hat{\mathbf{n}})\, \mathcal{U}   =\sigma_z$, to rotate our Hamiltonian towards the local  axis specified by $\hat{\bf{n}}$: 
\begin{equation}
H \to H' = \frac{ \left(
\boldsymbol{\pi} + e  \boldsymbol{\mathcal{A}}(\mathbf{X})
\right)^2
 }{2 \mass} +  \xc\ \sigma_z,
\end{equation}
where the potential $\boldsymbol{\mathcal{A}}$ comprises the mixing of two gauge fields:
$\boldsymbol{\mathcal{A}} = \mathcal{U}^\dagger \boldsymbol{\mathcal{A}}_\text{R} \,\mathcal{U}+ \boldsymbol{\mathcal{A}}^\text{ xc}$, with the additional contribution $\boldsymbol{\mathcal{A}}^\text{ xc} =- i\hbar \,\mathcal{U}^\dagger \nabla \mathcal{U} / e$.
The real-space Berry curvature corresponding to this vector potential can be recast as an effective, chirality- and spin-orbit-driven magnetic field $\mathcal{B}^{z}_{\text{eff}} = \sgn(\xc) \bra{\downarrow} \partial_{x} \mathcal{A}_y - \partial_{y} \mathcal{A}_x \ket{\downarrow}$ accompanying the ``ferromagnetic" system~\cite{Bliokh2005,Gorini2010,Fujita2011}:
\begin{equation}
\mathcal{B}^{z}_{\text{eff}} {\big |}_{\text{spiral}}
 = \frac{ m \soi q}{e}\sgn(\xc) \cos \theta
\end{equation}
Thus, in the limit of  
$|\soi| \ll |\xc|$, the emergence of chiral orbital magnetization can be understood as the coupling of a mixed $SU(2)$ gauge field to the diamagnetic Landau-Peierls susceptibility $\chi_{\text{LP}}^{\uparrow + \downarrow } = -e^2 / (12\pi m_{\text{e}})$, i.e., one can show that
\begin{equation}\label{eq:com2}
\com =  \frac{1}{2} \chi_{\text{LP}}^{\uparrow + \downarrow } \mathcal{B}_{\text{eff}}^z \
\sgn(\xc),
\end{equation}
in the vicinity of the band extrema.

\begin{figure}[t]
 \centering
 \includegraphics[width=0.9\linewidth]{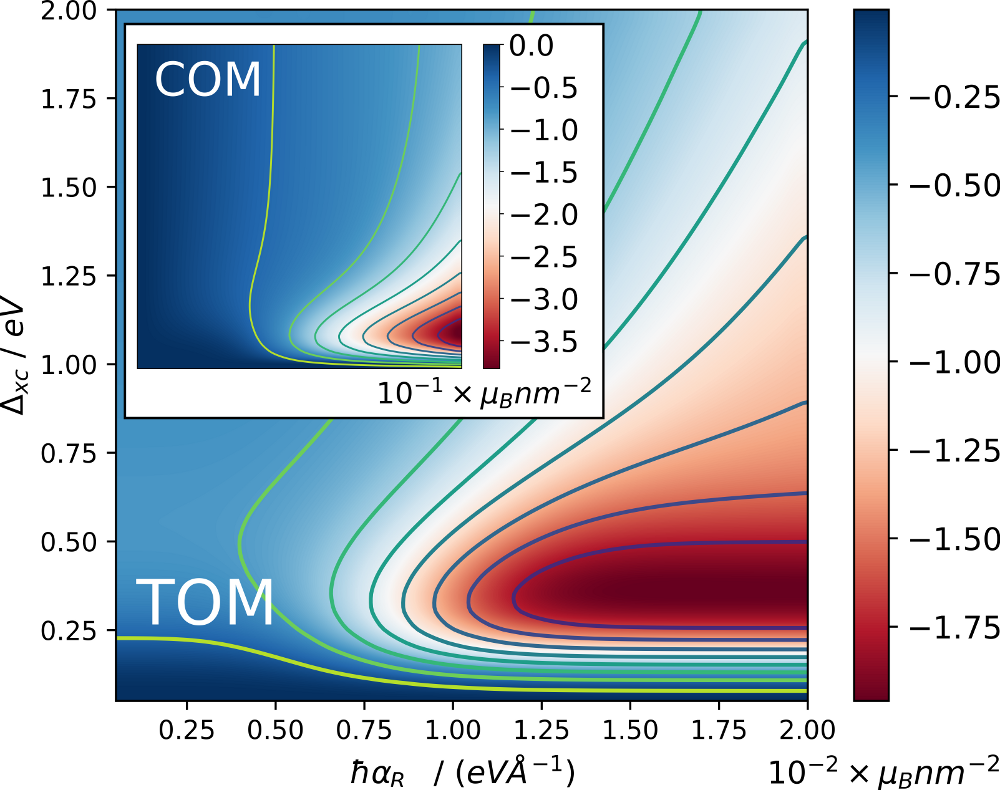}\\
 \caption{Phase diagram of $m_\text{tom}$, Eq.~(\ref{eq:tom}), evaluated at the core of a N\'eel-type skyrmion ($m=1$, $c=0.9$\,nm, $w=1.2$\,nm) as a function of the parameters $\xc$ and $\soi$ of the Rashba Hamiltonian (\ref{rashba:ham}) with $\mu=0$. Inset: phase diagram of $\com$, Eq.~(\ref{eq:com}), evaluated at the position of a spin-spiral ($q = 2.86$\,nm$^{-1}$) with an out-of-plane magnetization as a function of $\xc$ and $\soi$. 
% $\xc \gg \soi$, $\xc \gtrsim 0.5\ eV$ corresponds to the coupling of the emergent magnetic field to the diamagnetic Landau-Peierls susceptibilty. 
 In an intermediate regime of $\xc \lesssim \soi$ orbital magnetism is strongly enhanced. 
 %The inset shows the integrated TOM over the whole Skyrmion at $\xc = 0.9\ eV$ for three different stages of $\mu$.
 }\label{fig:phasediag}
\end{figure}

The behavior of COM becomes complicated and deviates remarkably from that given by Eq.~(\ref{com:model}) as the Rashba parameter increases. To demonstrate this, we numerically calculate the value of $\com$ for a spin-spiral with $q = - 2.86$\,nm$^{-1}$~\cite{Ferriani2008} at the position in real space with an out-of-plane magnetization, in a wide range of parameters
$\xc$ and $\soi$ of the Rashba Hamiltonian (\ref{rashba:ham}) with $\mu=0$, presenting the results in the inset of Fig.~\ref{fig:phasediag}. In this plot, we observe that while the gauge field picture is valid in the limit of $\xc / \soi \to \infty$, 
%and $\com$ disappears for $\soi / \xc \to \infty$
there exists a pronounced region in the ($\soi$,$\xc$)-phase-space
where COM exhibits a strong non-linear enhancement. In this region, the real-space behavior of $\com$
can be very non-trivial, deviating strongly from the $\cos$-like behavior of Eq.~(\ref{com:model}), see for 
example Fig.~\ref{fig:localdens}(a) for the distribution of COM along the spin-spiral for specific values of $\xc = 0.9$\,eV, $\hbar\soi = 2.0$\,\,eV$\mbox{\AA}^{-1}$, $k_\text{B} T = 0.1$\,eV. Such behavior can be related to the 
strongly anisotropic properties of the Rashba model, as discussed in detail below.

{\itshape Topological Orbital Magnetization.}~The TOM appears as the correction to the OM which is second order in the gradients of the texture, Eq.~(\ref{eq:tom}), and while it vanishes for one-dimensional spin-textures discussed above, we show that it is finite for 2D textures such as skyrmions. 
Remarkably, in contrast to COM, the TOM is non-vanishing even without spin-orbit interaction. To investigate this, we set $\soi$ to zero, reducing the effective vector potential to $\boldsymbol{\mathcal{A}} = \boldsymbol{\mathcal{A}}^\text{ xc}$ and with the emergent field turning into $B_\text{ eff}^z$, Eq.~(\ref{eq:Beff})~\cite{Fujita2011}. The gradient expansion now reveals that

\begin{equation}\label{eq:tom2}
\tom  = 
\frac{1}{4} \chi_{\text{LP}}^{\uparrow + \downarrow } B_{\text{eff}}^z \
\sgn(\xc) 
\left( 1 - 3\ \frac{\mu^2}{\xc^2} \right),
\end{equation}
if $ |\mu| < | \xc | $, and zero otherwise. 
In the vicinity of the band edges, where $|\mu| \approx |\xc|$, this again confirms the gauge-theoretical expectation, characterizing TOM as the electronic response to the emergent magnetic field (details on how the scalar spin chirality is entering this equation can be found in the supplement). Remarkably, the similarity between Eqs.~(\ref{eq:com2}) and (\ref{eq:tom2}) underlines the common origin of the COM and TOM in the ``effective" magnetic field in the system, generated by a combination of a gradient of $\hatn$
along $x$ with spin-orbit interaction (in case of COM), and by a combination of the gradients
of $\hatn$ along $x$ and $y$ (in case of TOM).

%PHASE-DIAGRAM
To explore the behavior of TOM in the presence of spin-orbit interaction, $\soi \neq 0$, we numerically compute the value of TOM at the center
of a N\'eel skyrmion of core size $c=0.9$\,nm, with the domain wall width $w = 1.2$\,nm and the topological charge $N_\text{sk} = -1$ (see Fig.~\ref{fig:localdens} and the supplement describing the exact modelling of the skyrmion shape as put forward in~\cite{Romming2015}), as function of $\xc$ and $\soi$ (at $\mu=0$). The corresponding phase diagram, presented in Fig.~\ref{fig:phasediag}, displays two notable features. The first one is the relative stability of Eq.~(\ref{eq:tom2}) against a perturbation by a spin-orbit field in the limit of $|\xc| \gg |\soi|$. The second one is the significant enhancement of TOM in the regime where $|\soi| >|\xc|$, similar to COM (albeit over a larger part of the parameter space). 
As exemplified in Fig.~\ref{fig:localdens}(b), the ``local" TOM in this regime of enhancement deviates strongly in its real-space distribution from the uniform behavior described by Eq.~(\ref{eq:tom2}) and is therefore not well described by $B_\text{ eff}^{z}$. This effect, again in direct analogy to COM, can partially be attributed to the enhanced diamagnetic susceptibility of the Rashba model~\cite{Schober2012}. And although replacing $\chi_{\text{LP}}$ in Eq.~(\ref{eq:tom2}) with the exact orbital magnetic susceptibility (as~e.g.\ given by the Fukuyama's result~\cite{Fukuyama1971}) can account for the behavior of the TOM around $\hatn = \pm \vec{e}_z$, it fails to reproduce the strongly anisotropic feature near $\hatn \perp \vec{e}_z$, prominent in Fig.~\ref{fig:localdens}(b).

%K-SPACE behavior

\begin{figure}
 \centering
 \includegraphics[width=0.9\linewidth]{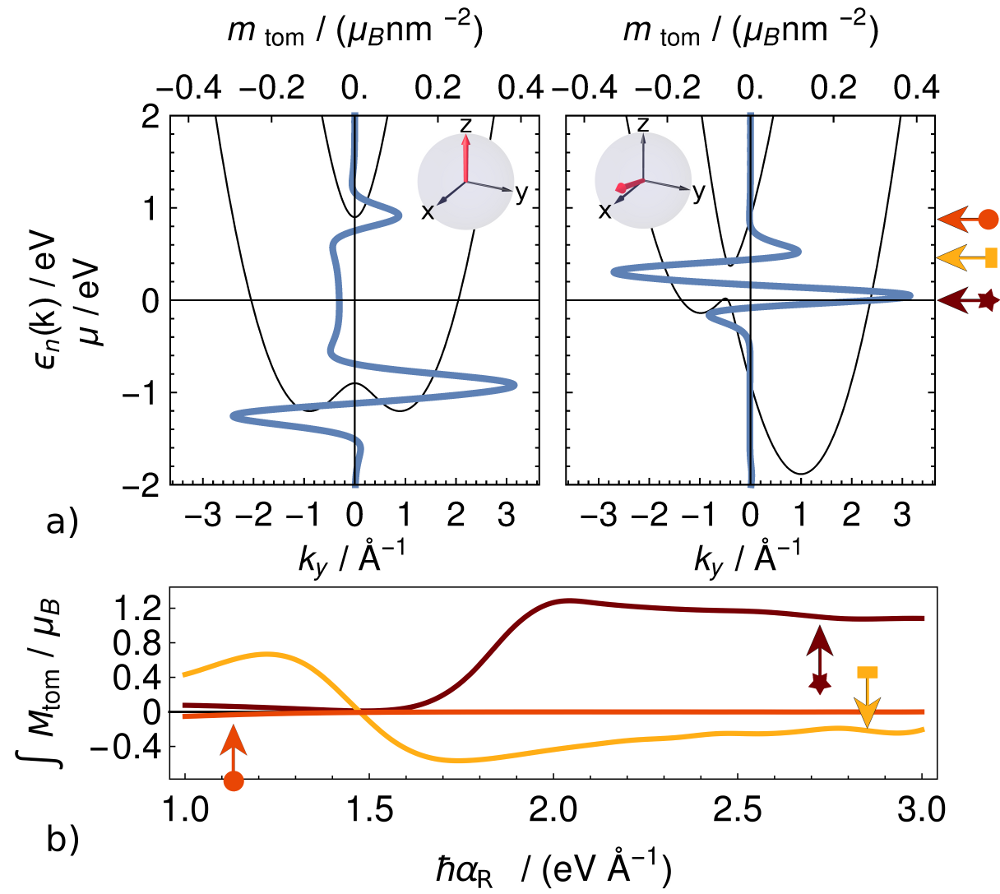}
 \caption{(a) TOM density (thick blue line) at two different positions (indicated by the red arrow) within the $N_\text{sk} = -1$ N\'eel Skyrmion  ($c=0.9$\,nm, $w = 1.2$\,nm, $\xc = 0.9$\,eV, $\soi = 2.0$\,eV\!$\mbox{\AA}^{-1}$) as a function of the chemical potential $\mu$ across the band structure $\epsilon_n$ (thin black lines) of the Rashba Hamiltonian~(\ref{rashba:ham}). The arrows
 mark the values of $\mu$ used in (b). (b) Integrated TOM (i.e., the topological orbital moment, $M_\text{tom}$) of the skyrmion with parameters from (a)  as a function of $\soi$ for different values of $\mu$. %\todo{Mistake: symbols wrong} 
 }
  \label{fig:muscan}
\end{figure}

This anisotropy is a direct consequence of the well-known non-trivial anisotropic $k$-space topology  of the 2D Rashba model~\cite{Shen2004}, 
which we can analyze by evaluating $\tom$ as function of the chemical potential $\mu$.
%and we next analyze the behavior of $\tom$ as a function of the chemical potential.
The results shown in Fig.~\ref{fig:muscan}(a) for $\hatn = \vec{e}_z$ reveal the sensitivity of $\tom$ to the $\soi$-induced deformation of the purely parabolic free-electron bands separated by $\xc$. The magnitude of TOM is largest and it exhibits pronounced oscillations in a narrow energy interval around the band edges. When we turn $\hatn$ into the in-plane direction, a $k\rightarrow-k$ asymmetric  $\soi$-driven band crossing occurs along $\mathbf{k}\perp\hatn$, 
eventually pushing the peaks of $\tom$ through the chemical potential which explains the anisotropy of TOM.
Besides a strong enhancement of TOM,
the effective real-space magnetic field picture, as given by Eq.~(\ref{eq:tom2}), fails to reproduce the strong
anisotropy of $\tom$, although an appropriate more general geometric theory, accounting for these features and formulated in terms of mixed Berry curvatures in real- and momentum-space, could exist~\cite{Freimuth2013}. 

%INTEGRATED VALUE
At this point, we turn to the discussion of the total integrated values of the orbital moments in chiral spin textures. As concerning the total value of the 
COM-driven orbital moment in one-dimensional 360$^\circ$ or 180$^\circ$ chiral domain walls 
it always vanishes identically by arguments of 
symmetry. In sharp contrast, the TOM-driven total orbital moment of skyrmions generally does not vanish. This can be first shown in the limit when the gauge-field approach is valid (i.e., ~$\xc  \gg \soi$). In this case, as follows from Eq.~(\ref{eq:tom2}),  
%in this limit
the integrated value of $\tom$ over the skyrmion, i.e., the topological orbital moment $m_\text{tom}$, is {\it quantized} to a universal value of $\mu_\text{ B} N_\text{sk} / 12$, at $\mu = 0$, independent of the parameters of the electronic structure. In this limit the skyrmion of $N_\text{sk}\neq 0$ thus behaves as an ensemble of $N_\text{sk}$ effective particles which occupy a macroscopic atomic orbital with associated universal value of the orbital angular momentum of $\mu_B/12$.   

In the other limit of $\soi > \xc$ the magnitude of $\tom$ can
be enhanced drastically with respect to this value. To show this, we calculate $\tom$ for N\'eel-type skyrmions with $N_\text{sk} = -1$ at a fixed value of $\xc=0.9$\,eV while varying $\soi$ for three different values of $\mu$, see Fig.~\ref{fig:muscan}(b). The presented data reveals an increase in $m_\text{tom}$
up to as much as 1\,$\mu_B$ upon increasing $\soi$, which can be attributed to the enhanced values of $\tom$ in the vicinity of the band extrema and its strong anisotropy in the considered part of the phase diagram, Fig.~\ref{fig:phasediag}. Another remarkable consequence of the non-trivial behavior of $\tom$ in energy, Fig.~\ref{fig:muscan}(a), is the dependence of the sign of $M_\text{tom}$ on the value of $\mu$. A fundamental result of our analysis is an almost complete independence of
the values of $m_\text{tom}$, discussed above, with respect to the parameters which determine the shape of the skyrmion, including its radius. Given the  observed remarkable  stability of $m_\text{tom}$ 
, we can truly call this moment {\it topologically-protected} in the sense of its robustness with respect to diverse perturbations of the underlying spin texture which gives rise to it. 

On a fundamental level, COM and TOM arise as a consequence of the changes in the local electronic structure in response to an emergent field.
This opens a way to experimentally access $\tom$ and $\com$ by such techniques as off-axis electron holography~\cite{Shibata2017} (sensitive to local distribution of magnetic moments), or scanning tunneling spectroscopy (sensitive to the local electronic structure) in terms of $B$-field induced changes in the $dI/dU$ or $d^2I/dU^2$ spectra~\cite{Kubetzka2017}.
%, or off-axis electron holography~\cite{X} (probing magnetic moments)
Further, the emergence of COM and TOM can give a thrust to the field of electron vortex beam microscopy~\cite{Fujita2017} $-$ where a beam of incident electrons intrinsically carries orbital angular momentum interacting with the magnetic system $-$ into the realm of chiral magnetic systems. For example, we speculate that at sufficient intensities, electron vortex beams can imprint skyrmionic textures possibly by partially transforming its orbital angular momentum into TOM.

At the end, we conclude that the phenomenon of chiral and topological orbital magnetism  marks outstanding prospects for ``chiral" spintronics and orbitronics. While the magnitude and details of the ``local" TOM and COM can be tuned by electronic structure engineering, the topological orbital moment is a new type of  
property
in the physics of skyrmions. This observable, in analogy to the topological charge, exhibits either a quantization or a strong protection against deformations of
the underlying spin structure, and thus provides unique means for skyrmion detection, manipulation and utilization. In particular, the topological orbital moment can be envisaged to mediate the interaction between skyrmions and circularly-polarized light, giving an opportunity for optical detection and control of the skyrmion topological charge. On the other hand, since the topological orbital moment is directly proportional to the topological charge of the skyrmions, we suggest that the interaction of TOM with external magnetic fields could be used to trigger the formation of skyrmions with large topological charge. Ultimately, the currents of skyrmions can be employed for ``lossless" transport of the associated topological orbital momenta over large distances in skyrmionic devices, opening new perspectives in orbitronics.

%ACKNOWLEDGEMENT

We thank J.-P. Hanke, M.d.S. Dias and S. Lounis for fruitful discussions, and gratefully acknowledge computing time on the supercomputers JUQUEEN and JURECA at Jülich Supercomputing Center, and at the JARA-HPC cluster of RWTH Aachen. We acknowledge funding under SPP 1538 and project MO 1731/5-1 of Deutsche Forschungsgemeinschaft (DFG) and the European Union's Horizon 2020 research and innovation programme under grant agreement number 665095 (FET-Open project MAGicSky).

\bibliography{letter}
\bibliographystyle{apsrev4-1}

\end{document}